\documentclass[asp,prl,amsmath,amssymb,reprint,superscriptaddress,preprintnumbers,showpacs,intlimits,shortbibliography]{revtex4-2}
\pdfoutput=1
\bibpunct{[}{]}{,}{n}{}{}

\usepackage[utf8]{inputenc}
\usepackage{bm,latexsym,mathrsfs,enumerate}
\usepackage[dvipsnames]{xcolor}
\usepackage{mathtools}
\usepackage{makecell}
\usepackage{multirow}
\usepackage[breaklinks=true,unicode=true,urlcolor = blue,colorlinks = true,citecolor = blue,linkcolor = blue]{hyperref}
\usepackage{graphicx}
\usepackage{ragged2e}
\usepackage[export]{adjustbox}
\usepackage[final]{pdfpages}
\usepackage{tikz}
\makeatletter
\AtBeginDocument{\let\LS@rot\@undefined}
\makeatother
\begin{document}

 \title{Magnons and fundamental magnetic interactions in a ferromagnetic monolayer: The case of Ni monolayer}

\author{Khalil~Zakeri}
\email{khalil.zakeri@partner.kit.edu}
\affiliation{Heisenberg Spin-dynamics Group, Physikalisches Institut, Karlsruhe Institute of Technology, Wolfgang-Gaede-Str. 1, D-76131 Karlsruhe, Germany}
\author{Albrecht~von~Faber}
\affiliation{Heisenberg Spin-dynamics Group, Physikalisches Institut, Karlsruhe Institute of Technology, Wolfgang-Gaede-Str. 1, D-76131 Karlsruhe, Germany}
\author{Arthur Ernst}
\affiliation{Institute for Theoretical Physics, Johannes Kepler University, Altenberger Strasse 69, A-4040 Linz,
	Austria}
\affiliation{Max-Planck-Institut f\"ur Mikrostrukturphysik, Weinberg 2, D-06120 Halle, Germany}
\begin{abstract}
\end{abstract}

%\date{\today}
\maketitle
\textbf{The experimental investigations of the magnetic interactions in an atomically thin magnetic layer are essential to understand the physics of low-dimensional magnets. The full spectrum of collective magnetic excitations (magnons) would provide an access to these fundamental  interactions on the atomic scale. Here in order to be able to excite the magnons by means of spin-polarized electrons we couple a Ni monolayer to one and two atomic layers of Co and probe the full experimental magnon dispersion relation up to the Brillouin zone boundary. Comparing to the results of \textit{ab initio} calculations we quantify the complex pattern of the magnetic exchange interaction in the Ni monolayer. We show that although the magnons in this system are rather stiff, the Heisenberg exchange coupling between the Ni spins is weak.  We unravel the origin of the observed large magnon stiffness constant being a consequence of the small spin density of the Ni atoms.}\\
The collective excitations of magnetic solids can be
described by their representative quasi-particles, called magnons. Of particular interest are the high wavevector magnons, since they are
governed by the magnetic exchange interaction between neighboring spins \cite{Heisenberg1928}. This interaction, also known as Heisenberg exchnage interaction (HEI), is indispensable for understanding many phenomena in magnetism \cite{Halilov1998,Pajda2001,Katsnelson2000}.
The Heisenberg spin Hamiltonian describing this fundamental interaction is usually given by $\mathcal{H_{\mathrm{HEI}}}=-\sum_{i\neq j}J_{ij}\mathbf{S}_i \cdot\mathbf{S}_j$, where the exchange  parameter $J_{ij}$, describes the interaction between atomic spins $\mathbf{S}_i$ and $\mathbf{S}_j$ sitting on sites $i$ and $j$. The pattern of HEI in low-dimensional itinerant magnets can be very complex \cite{Chuang2012,Meng2014,Prokop2009,Tang2007,Zakeri2013a, Meng2014}. Fortunately, probing the full magnon dispersion relation provides a direct and unambiguous way of resolving the complex pattern of HEI \cite{Zakeri2013a,Zakeri2014,Zakeri2017}.

Several experimental techniques based on the inelastic scattering of neutrons, photons and electrons have been successfully implemented to probe magnons in bulk and ultrathin films of Fe and Co down to the monolayer and even submonolayer regime. However, so far it has been challenging to probe the full magnon dispersion relation in Ni, in particular a Ni monolayer. This may have several  reasons. (i) The inelastic scattering cross-section of neutrons as bulk probes  by magnons scales with the magnetic form factor which, in turn, is directly related to the spin density of the unit cell. Owing to its rather small magnetic moment and its itinerant magnetism character, Ni has not been in the favor of the neutron scattering experiments \cite{Chatterji2006}. The magnon spectrum of bulk Ni has only been measured over a very small range of momentum \cite{Alperin1967,Mook1985,Mitchell1985}. (ii) Likewise, resonant inelastic x-ray scattering experiments have been limited by effects associated with the x-ray fluorescence \cite{Ament2011}. In the most recent experiments only a small fraction of the Brillouin zone (BZ) could be covered \cite{Brookes2020}.  (iii) Theoretical calculations of the inelastic electron scattering cross-section have revealed that due to the small exchange splitting and the presence of Stoner excitations at low energies it is practically not feasible to measure the magnon dispersion relation in Ni thin films and at Ni surfaces when using electrons as probing particles \cite{Hong2000}. All  these together have made the full magnon dispersion relation in Ni inaccessible. Given the fact that the neutron and x-ray scattering techniques do not have the monolayer sensitivity the experimental magnon spectrum as well as the magnetic exchange parameters in a Ni monolayer have remained hitherto fully unknown.
Meanwhile, however, the theory of magnetic excitations has been well advanced \cite{Halilov1998,Pajda2001,Karlsson2000,Sasioglu2010,Buczek2011,Rousseau2012,Mueller2016,Eriksson2016,Cao2018,Singh2019,Solovyev2021,Skovhus2021}.
In order to verify the validity of the theoretical approaches, the authors had to compare their results to the only one available set of the experimental data, which covers a very small fraction of the Brillouin zone \cite{Alperin1967,Mook1985,Mitchell1985}.

The magnons and the fundamental magnetic interactions in the Ni monolayer are also of great interest in the context of unconventional topological superconductivity in the Bi/Ni bilayer, where the interfacial Ni magnons are proposed to be responsible for the superconducting pairing mechanism being of  $d_{xy}\pm i d_{x^2+y^2}$ character \cite{Gong2017}.

Besides their importance for the fundamental understanding of the physics of low-dimensional solids, magnons are also of great interest to the field of magnonics, where the idea is to use the magnons as information carriers \cite{Kruglyak2010,Chumak2015,Zakeri2018,Zakeri2020}.

In this Letter by performing spin-polarized high resolution electron energy-loss spectroscopy (SPHREELS) experiments on specifically designing multilayer structures we probe, for the first time, the full dispersion relation of the magnon mode, which is partially localized in the Ni monolayer. We show that the Ni magnons are rather stiff and show a notable dispersion relation up to the surface BZ boundary. This is in sharp contrast to the results of the Fe monolayer on the same substrate or on W(110) and Pd(001), where the magnons are found to be rather soft. Comparing the results to those of first-principles calculations we shall comment on the observed large magnon stiffness constant. Moreover, we will provide the complex pattern of HEI in the Ni monolayer.

We have shown earlier that in multilayer structures the spatial localization of the magnons depends on the pattern of $J_{ij}$ \cite{Zakeri2013a,Zakeri2021}. Under some circumstances one may selectively excite magnons which are localized either predominantly at the  surface or interface, or almost equally at both. Moreover, previous investigations have revealed that the excitation cross-section of magnons with electrons is the highest when the sample surface is composed of Co atoms \cite{Vollmer2003,Etzkorn2005,Etzkorn2007,Ibach2014,Chen2017,Zakeri2021a,Zakeri2021}. Hence, in order to substantially enhance the magnon excitation cross-section we cover the Ni monolayer with one and two atomic layers of Co and designed the following multilayers: Co/Ni/Ir(001) and 2Co/Ni/Ir(001). A similar idea has recently been used to probe the standing spin waves with a parallel momentum $q_{\parallel}<0.3$~\AA$^{-1}$ in Co/Ni multiyers composed of several atomic layers  \cite{Ibach2018,Ibach2019}.

\begin{figure}[t!]
	\centering
	\includegraphics[width=.99\columnwidth]{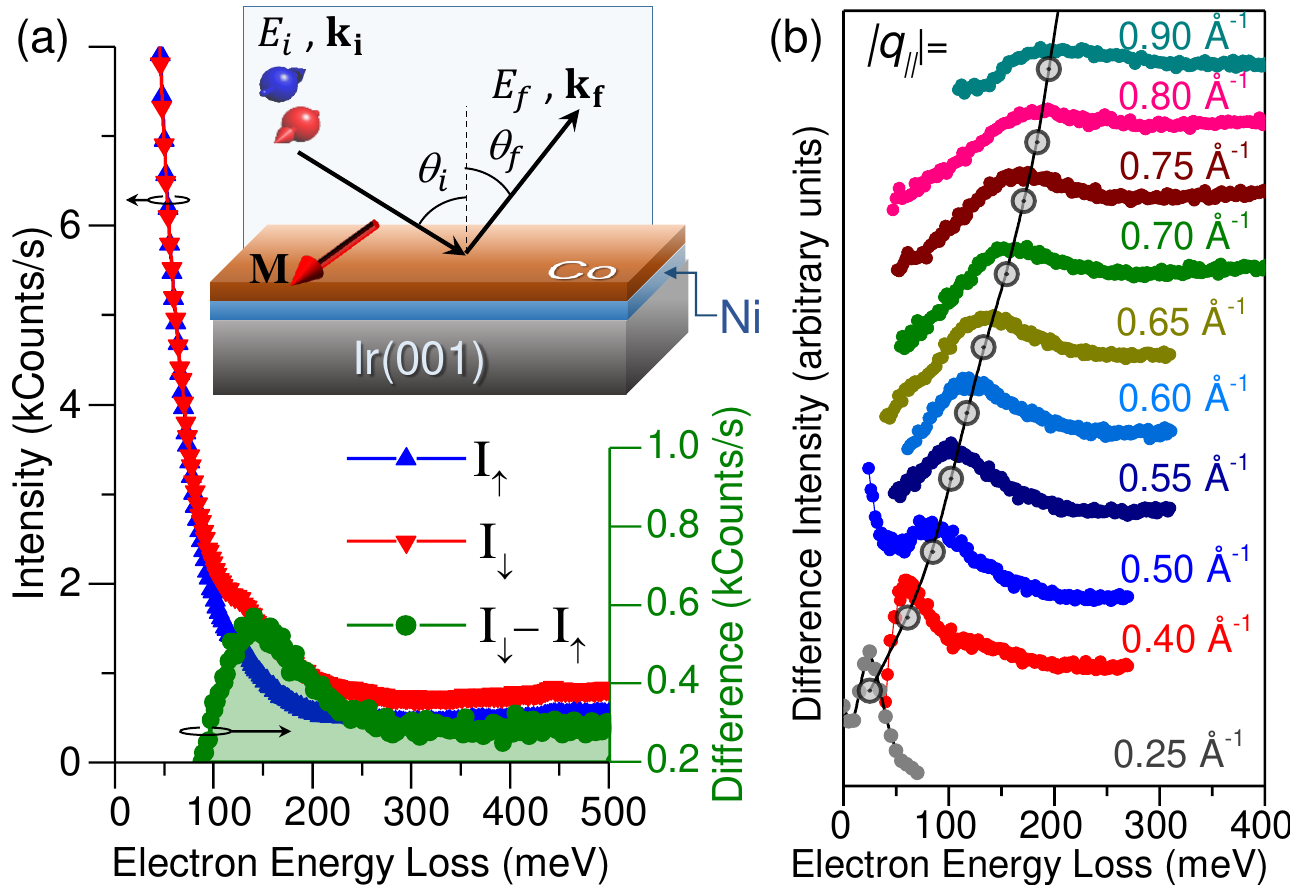}
	\caption{(a) Typical SPHREELS spectra recorded on the Co/Ni/Ir(001) structure at $|q_{\parallel}|=0.7$~\AA$^{-1}$. The spectra were recorded at the incident energy of $E_i=10$~eV and at room temperature. The red and blue spectra, denoted by $I_{\downarrow}$  and $I_{\uparrow}$, were recorded with the spin polarization vector of the incident electron beam being parallel and antiparallel to the magnetization $\mathbf{M}$, respectively. The difference spectrum $I_{\downarrow}-I_{\uparrow}$ is shown by the green color. The scattering geometry is schematically illustrated in the inset. The energy and wavevector of the incident (scattered) beam are shown by $E_i$ and $\mathbf{k_i}$  ($E_f$ and $\mathbf{k_f}$), respectively. (b) A series of difference spectra recorded for different magnon wavevectors $|q_{\parallel}|$ ranging from 0.25 to 0.90~\AA$^{-1}$. The excitation energy (peak position) is marked by the black circles.}
	\label{Fig1:Spectra}
\end{figure}

All the sample preparation and magnon spectroscopy experiments were performed under ultrahigh vacuum conditions.
We first examine the Co/Ni/Ir(001) epitaxial system. The surface of  Ir(001) was cleaned using the standard cleaning procedure, described in details in Refs. \cite{Zakeri2010a,Chuang2014,Zakeri2021}. The procedure leads to a well-ordered (1$\times$5) reconstructed surface. The Ni and Co monolayers were epitaxially grown by molecular beam epitaxy at room temperature. Figure~\ref{Fig1:Spectra}(a) shows typical SPHREELS spectra recorded on the Co/Ni/Ir(001) structure at a wavevector of $|q_{\parallel}|=|\Delta k_{\parallel}|=|k_{i \parallel}-k_{f \parallel}|=0.7$~\AA$^{-1}$, where $k_{i \parallel}$ and $k_{f \parallel}$ denote the parallel momentum of the incoming and outgoing beam, respectively [see the inset of Fig.~\ref{Fig1:Spectra}(a)]. The spectra were recorded for the two possible spin orientations of the incoming electron beam, i.e., parallel ($I_{\downarrow}$, red color) and antiparallel ($I_{\uparrow}$, blue) to the sample magnetization $\mathbf{M}$. Due to the conservation of the total angular momentum during the scattering event, the magnons are only excited by incidence of minority electrons (electrons with their spin parallel to $\mathbf{M}$). Hence, the difference spectrum $I_{\downarrow}-I_{\uparrow}$ includes all the information regarding the magnons excitation energy (or frequency) \cite{Zakeri2013, Zakeri2014b} and lifetime  \cite{Zhang2011, Zakeri2012,Zhang2012}. In this experiment $\mathbf{M}$ was parallel to the $[\overline{1}10]$-direction and the magnon wavevector $q_{\parallel}$ was along the $[110]$-direction. This corresponds to the $\bar{\Gamma}$--$\bar{\rm X}$ of the surface BZ.
The magnon dispersion relation was probed along this symmetry direction. Different magnon wavevectors were achieved by changing the scattering angles $\theta_i$ and $\theta_f$. A series of difference spectra recorded for different values of $|q_{\parallel}|$ ranging from 0.25 to 0.90 \AA$^{-1}$ is shown in Fig.~\ref{Fig1:Spectra}(b). The results clearly indicate that the measured magnon mode exhibits a rather stiff dispersion, similar to the standing spin waves in thick ferromagnetic layers \cite{Balashov2014}.
\begin{figure}[t!]
	\centering
	\includegraphics[width=0.99\columnwidth]{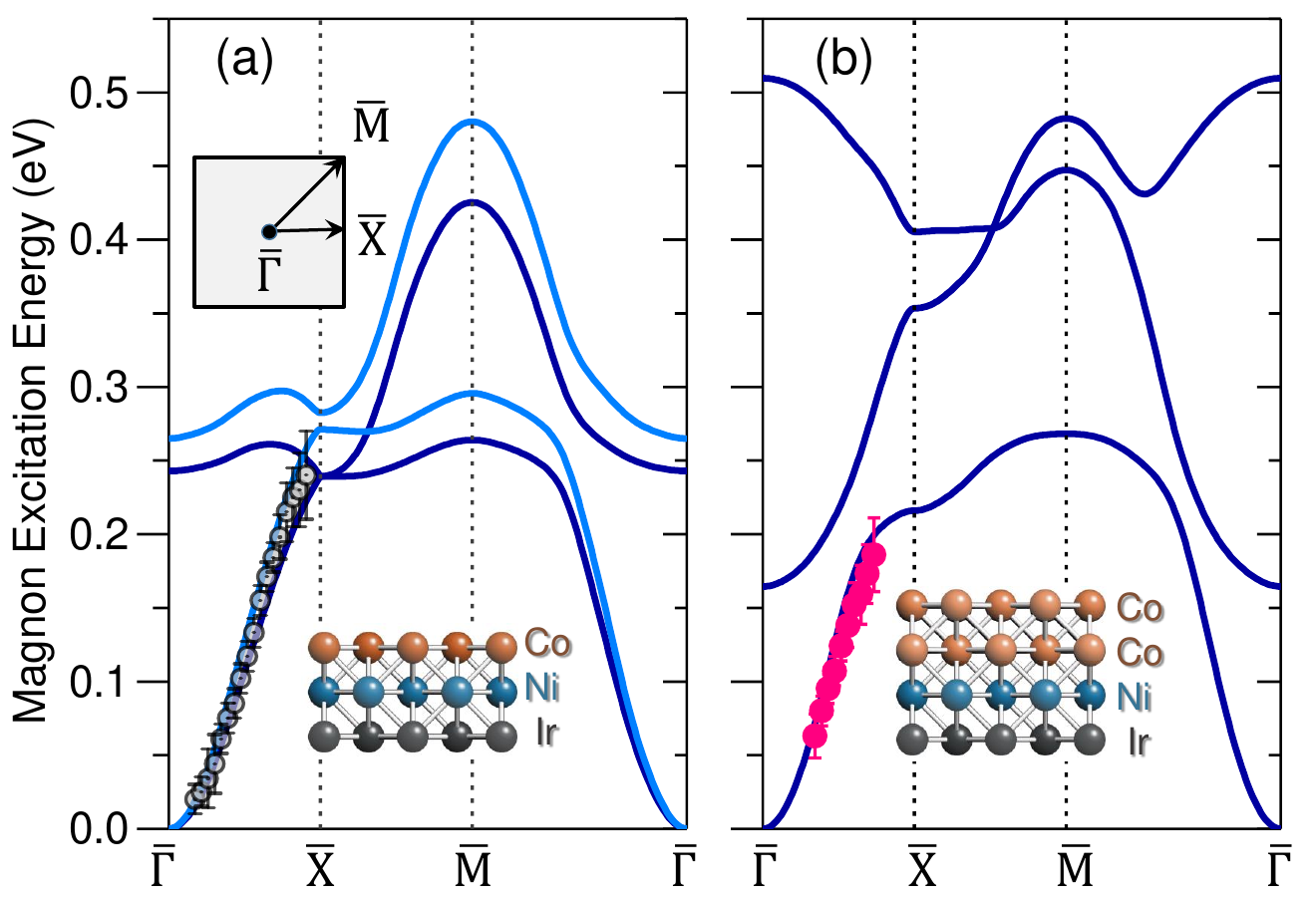}
	\caption{The magnon dispersion relation of (a) Co/Ni/Ir(001) and (b) 2Co/Ni/Ir(001). The calculated magnon dispersion relation is shown for two cases:  without (light-blue) and with (dark-blue)   considering the  reconstruction of the Ir surface. The lower insets show a schematic representation of the structures. The upper inset in (a) shows the surface BZ.}
	\label{Fig2:Dispersion}
\end{figure}
The resulting magnon dispersion relation is summarized in Fig.~\ref{Fig2:Dispersion}(a), representing the the acoustic magnon mode of the system. In order to experimentally verify the partial localization of this mode in the Ni monolayer and to ensure that the dispersion relation is governed by the exchange parameters in both Ni and Co layer, we added another Co monolayer on top of the structure and probed, once more, the magnon dispersion relation. The results of the 2Co/Ni/Ir(001) multilayer are shown in Fig.~\ref{Fig2:Dispersion}(b). The very similar dispersion relation for the two systems is an indication that the probed magnon mode is partially localized in the Ni monolayer.
However, comparing the results to those of Fe monolayer on W(110) \cite{Costa2008,Prokop2009,Bergman2010,Zakeri2014a}  and Pd(001) \cite{Qin2015,Qin2017}, or those of buried Fe monolayer on the same substrate \cite{Qin2019,Zakeri2021},  one realizes that in the present case the magnon dispersion relation is very stiff. This is surprising, since it has been found theoretically that both HEI and the Curie-temperature of Ni are much smaller than those of Fe and Co \cite{Halilov1998,Pajda2001,Karlsson2000,Solovyev2021}.

\begin{figure}[t!]
	\centering
	\includegraphics[width=0.99\columnwidth]{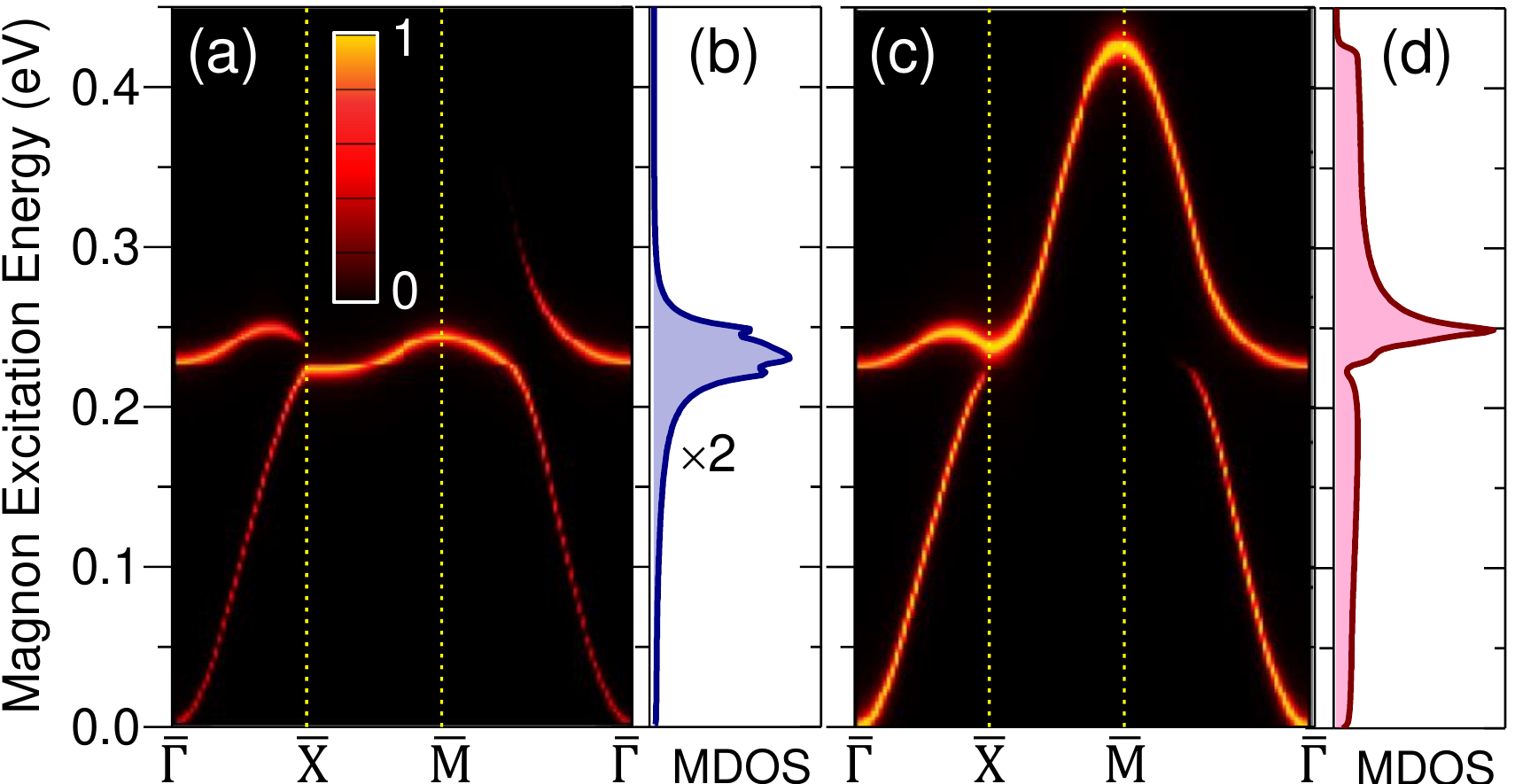}
	\caption{The magnon Bloch spectral function (BSF) and the magnon density of states (DOS) for the Co/Ni/Ir(001) system, projected onto the Ni monolayer [(a) and (b)] and onto the Co layer [(c) and (d)]. For the sake of clarity, the data in (b) are multiplied by a factor of 2.}
	\label{Fig3:MagBSF}
\end{figure}

In order to quantify the strength of HEI and to unravel the origin of the probed magnon mode we resort to the first-principles calculations of the magnetic exchange parameters in both Co/Ni/Ir(001) and 2Co/Ni/Ir(001) systems. Our first-principles calculations are based on density functional theory and the fully relativistic Korringa-Kohn-Rostoker electronic structure method. In this approach the values of $J_{ij}$ are computed within the framework of the magnetic force theorem \cite{Liechtenstein1987}. The experimental lattice parameters were used as the input of the first-principles calculations. The calculations provide most of the magnetic parameters of the systems including the matrix of $J_{ij}$. The magnon dispersion relation was then computed based on the $J_{ij}$-matrix. The results of calculations for the two systems are summarized in Fig.~\ref{Fig2:Dispersion}. The calculated magnon dispersion relation agrees well with the experimental results. Moreover, similar to the experiment the calculations indicate that the acoustic magnon mode of the two systems is very similar.
To unravel the origin of different magnon modes we also calculated the magnon Bloch spectral function (BSF) and the magnon density of states (DOS). When these quantities are projected onto different layers they would provide an access to the spatial localization of different magnon modes. In Fig.~\ref{Fig3:MagBSF} we provide the magnon  BSF and DOS for the Co/Ni/Ir(001) structure (the results of 2Co/Ni/I(001) are presented in Supplemental Figure~S1 \cite{Note1}). The projected BSF of  the two magnon modes, shown in Fig.~\ref{Fig3:MagBSF}, indicates that the spectral weight of the acoustic magnon mode near the high symmetry $\bar{\rm X}$ and $\bar{\rm M}$ points as well as along the $\bar{\rm X}$--$\bar{\rm M}$ path is the highest when magnon BSF is projected onto the Ni monolayer. This means that in this part of BZ this mode is almost entirely localized in the Ni layer. Likewise, the partial magnon DOS of the Ni monolayer [Fig.\ref{Fig3:MagBSF}(b)] is maximum in this range of energy. While above a magnon energy of about 250~meV both the magnon BSF and DOS are zero,  the low energy part of the magnon spectrum shows a finite spectral weight in the Ni monolayer. The projected magnon BSF and DOS onto the Co layer indicate that except for the high-symmetry $\bar{\rm X}$ and $\bar{\rm M}$ points and along the $\bar{\rm X}$--$\bar{\rm M}$ direction, the magnons have a considerable spectral weight in the Co layer. The main conclusion of the results presented in Fig.~\ref{Fig3:MagBSF} is that the experimentally probed magnon mode describes magnons, which have a finite spectral weight in both layers. The larger spectral weight of this mode in the Co layer makes it easily accessible to the electrons as probing particles. Hence, this mode is very efficiently excited in this system. A similar conclusion can be drawn for the 2Co/Ni/Ir(001) system \cite{Note1}.

\begin{table*}[t!]
	\centering
	\caption{The magnetic exchange parameters (in meV) and the magnetic moments (in $\mu_B$) as calculated by our first-principles calculations. Co$_1$ and Co$_2$ refer to the first and second Co atomic layer on top of the Ni monolayer. For the sake of simplicity only the HEI in the Ni monolayer up to the third nearest neighbors (NNN) are presented.}
	\label{Tab:HEI}
	\begin{tabular}{rcccccccccccc}
		\hline
		\hline
		Multilayer system & $J^{NiNi}_{N\parallel}$ & $J^{NiCo_1}_{N\perp}$  &$J^{NiNi}_{NN\parallel}$ & $J^{NiNi}_{NNN\parallel}$ & $J^{NiCo_1}_{NN\perp}$ & $J^{NiCo_1}_{NNN\perp}$  & $J^{NiCo_2}_{N\perp}$ & $J^{NiCo_2}_{NN\perp}$ & $J^{NiCo_2}_{NNN\perp}$&  $\mu_{Ni}$& $\mu_{Co_1}$&$\mu_{Co_2}$\\
		\hline
		
		Co/Ni/Ir(001)--1$\times$1  & 1.24 & 6.48 & 0.29   & -0.047 & 0.2  & -0.22  & -- & -- &  -- & 0.52 & 1.98 & --\\
		Co/Ni/Ir(001)--1$\times$5  & 1.03 & 5.71 & 0.164 & 0.002  & 0.24 & -0.25 &  -- &  -- & -- & 0.55 & 1.95 & --\\
		2Co/Ni/Ir(001)--1$\times$1& 2.75 & 6.28 & -0.001& 0.275  & 0.6   & -0.21 & -1.13 &  0.11 & -1.06 & 0.57 & 1.84 & 1.87 \\
		2Co/Ni/Ir(001)--1$\times$5& 1.84 & 4.93 & 0.008 & 0.178  & 0.55 & -0.21  & -0.69 & 0.24 & -0.79  & 0.56 & 1.81 & 1.88 \\
		\hline
		\hline
	\end{tabular}
\end{table*}
Our first-principles calculations indicate that the nearest neighbor intralayer and interlayer interaction with (without) considering the surface reconstruction are only $J_{N\parallel} = 1.03$ (1.24) meV and $J_{N\perp} = 5.71$ (6.48) meV, respectively. Likewise, the interaction between spins located at larger distances are also rather weak.  The second nearest neighbor intralayer and interlayer interaction with (without) considering the surface reconstruction are only  $J_{NN\parallel} = 0.164$ (0.29) meV and  $J_{NN\perp} = 0.24$ (0.2) meV, respectively. The values of $J_{ij}$ and the magnetic moments are provided in Tab.~\ref{Tab:HEI}. The fact that the HEI in Ni is weak is in agreement with the calculations for bulk fcc Ni \cite{Karlsson2000,Buczek2011,Rousseau2012,Mueller2016,Cao2018,Singh2019,Solovyev2021}. Note that due to the lack of the full magnon spectrum, the exchnage parameters in Ni are not known experimentally. The main conclusion of the results presented in Tab.~\ref{Tab:HEI} is that the exchange parameters in the Ni monolayer are rather small. Moreover, the pattern of the exchange interaction is complex and includes both ferromagnetic (positive) and antiferromagnetic (negative) exchange constants.
Owing to the weak HEI in the Ni monolayer one would expect a larger magnon BSF and DOS of the acoustic magnon mode in this layer. However, it is important to consider that the magnetic moment of Ni is by a factor of about 3.5 smaller than that of the Co. This means that the dynamic component of the magnetization (which somewhat scales with the static magnetic moment) is smaller in the Ni layer. Hence, one observes a larger magnon BSF and DOS when they are projected onto the Co layer. Note that the magnon BSF and DOS exhibit finite values in the Ni monolayer. The large magnon BSF and DOS in the Co layer helps to easily excite this magnon mode.

Near the $\bar{\Gamma}$--point the magnon dispersion relation may be approximated by $\mathcal{E}(q_{\parallel}) \simeq Dq_{\parallel}^2$, where $D$ denotes the magnon exchange stiffness constant. Fitting the data for $q_{\parallel}<0.4$~\AA$^{-1}$, we find a value of $356\pm5$~meV\AA$^2$, in reasonable agreement with the experimental  \cite{Alperin1967,Mitchell1985,Mook1985} and theoretical \cite{Karlsson2000,Buczek2011,Rousseau2012,Mueller2016,Cao2018,Singh2019,Solovyev2021} values of the bulk Ni. Due to the weak HEI in the Ni monolayer, this is somewhat surprising.
In the following we shed light on the origin of the stiff magnon mode in the Ni monolayer.

Within the adiabatic formalism (linear spin-wave theory) the magnon dispersion relation is given by $\mathcal{E}(q)= \frac{2g\mu_B}{\mu_i}\Sigma_{j\neq 0}J_{0,j}\left[1-\exp i(\mathbf{q}\cdot\mathbf{R}_{0j})\right]$, where $g=2$ is the  $g$-factor, $\mu_B$ is the Bohr magneton, $\mu_i$ is the magnetic moment of the origin site, $\mathbf{R}_{0j}$ represents the displacement vector of site $j$ with respect to the origin and $\mathbf{q}$ denotes the three dimensional vector of the magnon momentum. A careful investigation of the electronic DOS reveals that the Ni atoms possess a low spin density. A low spin density leads to both a weak exchange interaction as well as a small magnetic moment [0.55 (0.52)  $\mu_B$ for the reconstructed (unreconstructed) Ir(001)]. The fact that the magnetic moment of Ni is much smaller than that of the Co has  experimentally been verified by probing both bulk samples \cite{Billas1994} as well Co/Ni multilayers \cite{Wilhelm1999,Vaz2007,Andrieu2018}. Since the exchnage stiffens scales inversely with $\mu_i$, a weak HEI in Ni can lead to a large magnon exchange stiffness, comparable to that of the Co and Fe. Hence, assuming the same exchange constants for Ni and Co is not valid, even though for the small magnon momentum the dispersion relation of the two systems might be very similar \cite{Ibach2018,Ibach2019}.

Recently, the emergence of topological superconductivity in the Bi/Ni bilayer is attributed to the magnetic excitations at the interface and their coupling to the surface state electrons \cite{Gong2017}. We anticipate that our quantitative results on the values of HEI in the Ni monolayer would contribute to a quantitative understanding of superconducting mechanism in this system. Moreover, they would provide guidelines for tuning superconductivity in similar bilayer structures in which the magnons play a decisive role in the pairing mechanism \cite{Hugdal2018,Kargarian2016}.

In conclusion, we prepared atomically architectured  multilayers composed of Ni and Co  epitaxial monolayers on Ir(001). Owing to the large magnetic moment of the surface Co monolayer and the fact that the excitation cross-section at Co surfaces is high,  the magnons in such designed multilayers can be very efficiently excited by spin-polarized electron scattering experiments, e.g., SPHREELS. The acoustic magnon mode of these systems was found to be localized in both the Co as well as in the Ni monolayers. The full experimental acoustic magnon dispersion relation, probed up to the surface BZ, enabled us to quantitatively resolve the complex pattern of HEI in the Ni monolayer. We observed a rather weak exchange coupling within the Ni monolayer, even though the magnon exchange stiffness is rather large (much larger than that of an Fe monolayer on various substrates). The large magnon stiffness constant is a consequence of the low spin density of Ni atoms, as confirmed by our first-principles calculations. In addition to the fact that our results resolve the long-standing question regarding the quantitative values of HEI in the Ni monolayer, they are also of the interest to the field of topological superconductivity in magnetic/topological materials heterostructures.
\section*{Acknowledgments}

Financial support by the Deutsche
Forschungsgemeinschaft (DFG) through the DFG
Grants No. ZA~902/7-1 and No. ZA~902/8-1 is acknowledged. Kh.Z. thanks the
Physikalisches Institut for hosting the group and providing
the necessary infrastructure.
%\bibliographystyle {apsrev}
%\bibliography

\end{document}